\documentclass[twocolumn,10pt,superscriptaddress,amsmath,amssymb,aps,prb]{revtex4-2}

\pdfoutput=1

\usepackage[utf8]{inputenc}

\usepackage{amsmath}
\usepackage{amsfonts}
\usepackage{amsthm}
\usepackage{mathtools} % for dcases

\usepackage[hyperindex,breaklinks]{hyperref}

\usepackage[paperwidth=210mm,paperheight=297mm,centering,hmargin=2.5cm,vmargin=2.5cm]{geometry}

\usepackage{gensymb}
\usepackage{graphicx}
\usepackage{dcolumn}
\usepackage{array}
\usepackage{bm}% bold math
\usepackage{upgreek}
\usepackage{fancyhdr}
\usepackage{textcase}
\newcommand{\myFig}[6]{ %
\begin{figure}[htb]
\begin{center}
\includegraphics[width=#1\columnwidth,height=#2\columnwidth,clip=true,keepaspectratio=#3]{#4}
\caption{#5} \vspace{-0.5cm} \label{#6} 
\end{center}
\end{figure}}

\newcommand{\myFigWide}[6]{ %
\begin{figure*}[htb]
\begin{center}
\includegraphics[width=#1\columnwidth,height=#2\columnwidth,clip=true,keepaspectratio=#3]{#4}
\caption{#5} \vspace{-0.5cm} \label{#6} 
\end{center}
\end{figure*}}

\begin{document}

\preprint{AIP/123-QED}

\title{Ab initio comparison of spin-transport properties in MgO-spaced ferrimagnetic tunnel junctions based on Mn$_3$Ga and Mn$_3$Al}
% Force line breaks with \\
\author{M. Stamenova} \email[Contact email address: ]{stamenom@tcd.ie}
 \affiliation{School of Physics and CRANN, Trinity College Dublin, Dublin 2, Ireland.}%Lines break automatically or can be forced with \\

\author{P. Stamenov}
 \affiliation{School of Physics and CRANN, Trinity College Dublin, Dublin 2, Ireland.}%Lines break automatically or can be forced with \\

\author{N. Baadji}%
\affiliation{ Laboratoire de Physique des Matériaux et ses applications \& Département de Physique, Faculté des Sciences, Université Mohamed Boudiaf, M’sila, 28000, Algérie}
%

%\date{\today}% It is always \today, today,
             %  but any date may be explicitly specified

\begin{abstract}
We report on first-principles spin-polarised quantum transport calculations (from NEGF+DFT) in MgO-spaced magnetic tunnel junctions (MTJs) based on two different Mn-based Heusler ferrimagnetic metals, namely Mn$_3$Al and Mn$_3$Ga in their tetragonal DO$_{22}$ phase. The former is a fully compensated half-metallic ferrimagnet, while the latter is a low-moment high-spin-polarisation ferrimagnet, both with a small lattice mismatch from MgO. In identical symmetric and asymmetric interface reconstructions across a 3-monolayer thick MgO barrier for both ferrimagets, the linear response (low-voltage) spin-transfer torque (STT) and tunneling magneto-resistance (TMR) effects are evaluated. A larger staggered in-plane STT is found in the Mn$_3$Ga case, while the STT in Mn$_3$Al vanishes quickly away from the interface (similarly to STT in ferromagnetic MTJs). The roles are reversed for the TMR, which is practically 100\% in the half-metallic Mn$_3$Al-based MTJs (using the conservative definition) as opposed to 60\% in the Mn$_3$Ga case. The weak dependence on the exact interface reconstruction would suggest Mn$_3$Ga-Mn$_3$Al solid solutions as a possible route towards optimal trade-off of STT and TMR in the low-bias, low-temperature transport regime.
\end{abstract}

\maketitle

\section{\label{sec:level1}Introduction}

Among the Mn-based Heusler ferrimagnets are a number of topical binary materials, combining low moments with high Curie temperature and spin-polarisation \cite{Rode2013,Winterlik2008,Gao2013,Fan2020}, as well as high anisotropy with low Gilbert damping \cite{Mizukami2016,Mizukami2016_2} -- hence holding promise for the emerging field of Antiferromagnetic Spintronics\cite{Sinova2016,Jungwirth2016}. Ferrimagnetic electrodes, and especially, the compensated ones, in magnetic-tunnel junction (MTJ) devices lead to the reduction of the demagnetizing field and, more importantly, the reduction of the critical current needed to switch the magnetization together with the fast dynamics of such switching, involving inter-sublattice exchange interactions. The possible exploitation of this class of materials, however, depends on the ability to produce high-frequency oscillations (in the 100s of GHz range) and the depth of resistance modulation that junctions using these could support. Predictive computational guidance for the magnitude and resilience of both the spin-transfer torque (STT) and the tunneling magneto-resistance (TMR) effects, towards the difficult-to-control experimentally barrier quality and interface reconstruction, can help to speed-up the development of functional prototype devices. 

One such candidate is Mn$_3$Ga, which in its tetragonal DO$_{22}$ phase is a low-moment ferrimagnet with a high spin-polarisation and anisotropy \cite{Rode2013, Kharel2014, Winterlik2008}, but also a low Gilbert damping and an established epitaxial relationship with MgO(001) \cite{Mizukami2011}. A large staggered long-ranged STT effect has been found theoretically in MTJs based on DO$_{22}$-phased Mn$_3$Ga \cite{Stamenova2021}, present both in Fe/MgO/Mn$_3$Ga and in Mn$_3$Ga/MgO/Mn$_3$Ga tri-layers, and related to the mismatch of the Fermi wavevectors of the majority and minority $\Delta_1$ symmetry band in Mn$_3$Ga in the direction of transport. Theoretically, the TMR effect in these stacks reaches a few tens or percent and exhibits a sign change below 1V, which is in accordance with experimental observations for similar ferrimagnetic MTJs \cite{Borisov2016}.  

Another Mn-based Heusler Mn$_3$Al has been shown to exhibit half-metallicity and almost ideally fully compensated moment in its cubic DO$_3$ phase \cite{Jamer2017}, similarly to Mn$_3$Ga in this phase\cite{Gao2013}, and proposed MTJs with GaAs have shown large theoretical TMR ratios \cite{Qiu2021}. A GGA-PBE geometry optimisation of Mn$_3$Al reveals a stable tetragonal DO$_{22}$ solution with almost fully compensated moment and an in-plane lattice constant commensurate with MgO. As geometrically the Mn$_3$Ga and Mn$_3$Al stacks with MgO are very similar, but offer different placement of the Fermi level with respect to the main band dispersions, a comparison of the spin-transport in analogous MTJs could unveil further insights about the Spintronic capacity of the two Heuslers. Here we first examine the electronic structure properties of both materials in bulk (Section \ref{Sect1}) and then we compare the spin-dependent transport properties of two pairs of MTJs, all with 3-monolayers (ML) thick MgO spacers, but featuring two different terminations at one of the interfaces (Section \ref{Sect2}).     

\section{Bulk properties}\label{Sect1}

\myFigWide{1.8}{1.2}{true}{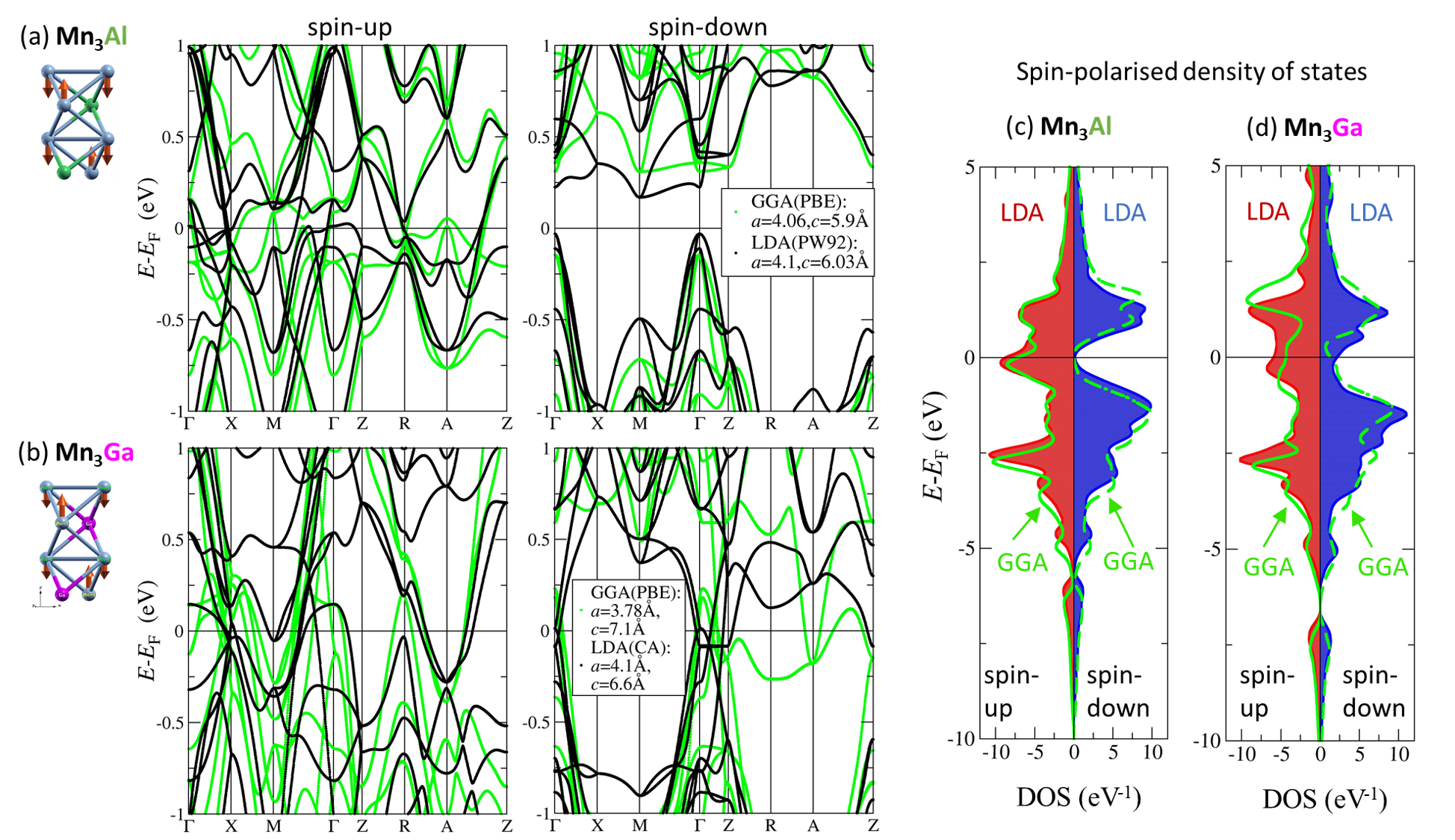}{ Band structures and spin-polarised density of states of (a,c) Mn$_3$Al and (b,d) Mn$_3$Ga, respectively in their depicted unit cell, calculated using LDA-PW92 (for Mn$_3$Al) and LDA-CA (for Mn$_3$Ga) on the MgO-matching (strained) geometries (black or red/blue filled curves, see text and insets). In all panels a comparison is shown with a corresponding GGA-PBE calculation for the relaxed unit cell (green curves, see text/legends for details). In (a,b) the DO$_{22}$ unit cells of both materials with the local spins of the Mn atoms shown as arrows.}{fig01}

Intermetallic Heusler alloys X$_2$YZ crystallize usually in the cubic L2$_1$ structure, especially at high temperature, and at low temperature they can develop a DO$_{22}$ tetragonal structure with a ratio $c/a$ around $\sqrt{2}$. The larger departure of this ratio from $\sqrt{2}$ is a prelude for a higher magneto-crystalline anisotropy \cite{Rode2013}. Our calculated optimum lattice parameters, using the GGA-PBE, of both compounds in their antiferromagnetic configuration are $a=4.057$, $c=5.911$~\AA~ and $a=3.78$, $c=7.1$~\AA~ for Mn$_3$Al and Mn$_3$Ga, respectively. The in-plane lattice constants are close to that of bulk MgO ($a_\mathrm{MgO}=4.21$~\AA), which makes their integration in conventional magnetic tunnel junctions feasible. The DO$_{22}$ Mn$_3$Al(Ga) structures are constructed by alternating planes of Mn-Al(Ga) (Mn in $2b$ Wyckoff position: Mn$_\mathrm{I}$) and Mn-Mn (Mn in $4d$ Wyckoff position: Mn$_\mathrm{II}$) coupled anti-ferromagnetically, along the $z$-axis (see unit cell schematics in Fig. \ref{fig01}). In such a lattice structure, Mn$_3$Ga is metallic and both spins contribute to the conductivity, while Mn$_3$Al is a half-metal and only spin-up states contribute to the conduction. Consequently, Mn$_3$Al has a 100\% spin polarization at the Fermi level and therefore one can expect a high TMR for junctions based on Mn$_3$Al. Mn$_3$Ga also shows a high spin polarization of 88\% in other GGA-based calculations \cite{Winterlik2008}. Similar results for the Fermi level spin-polarisation are obtained also when the transverse lattice constants are $a=b=4.1$~\AA~in both materials (approaching that of MgO thin films), as shown in the density of states presented in Fig. \ref{fig01}(c,d). 

The corresponding band structure of both compounds is presented in Fig. \ref{fig01}(a,b) and one can see that for Mn$_3$Al the spin-down channel exhibits a gap of the order of 0.4~eV when computed with GGA-PBE on the relaxed unit cell (green points) and similarly about 0.25~eV, when calculated with LDA-PW92 (the black points), as implemented in the {\sc Siesta} code \cite{Soler2002}. This gap can be tuned by changing the lattice parameters. In the LDA case, we apply a longitudinal tensile strain of 4\% ($c=6.027$~\AA) to open the gap and approximate the GGA result (where $c=5.795$~\AA). The impact of that on the layer-resolved magnetic moments (calculated by Mulliken population analysis) is a small decrease by about 7~\% for both Mn sublattices compared to GGA. We have additionally established that Mn$_3$Al keeps its fully-compensated ferrimagnetic character for applied strains ranging between -4\% to 8\% (so that the 100\% polarization is kept for such a strain). Note that the orbitals below the Fermi level are mainly a hybridisation between d-$t_{2g}$ orbitals of the Mn atoms, while the empty bands above the Fermi level are $e_g$ bands of the Mn occupying site $4d$. 

\myFigWide{2.5}{1.2}{true}{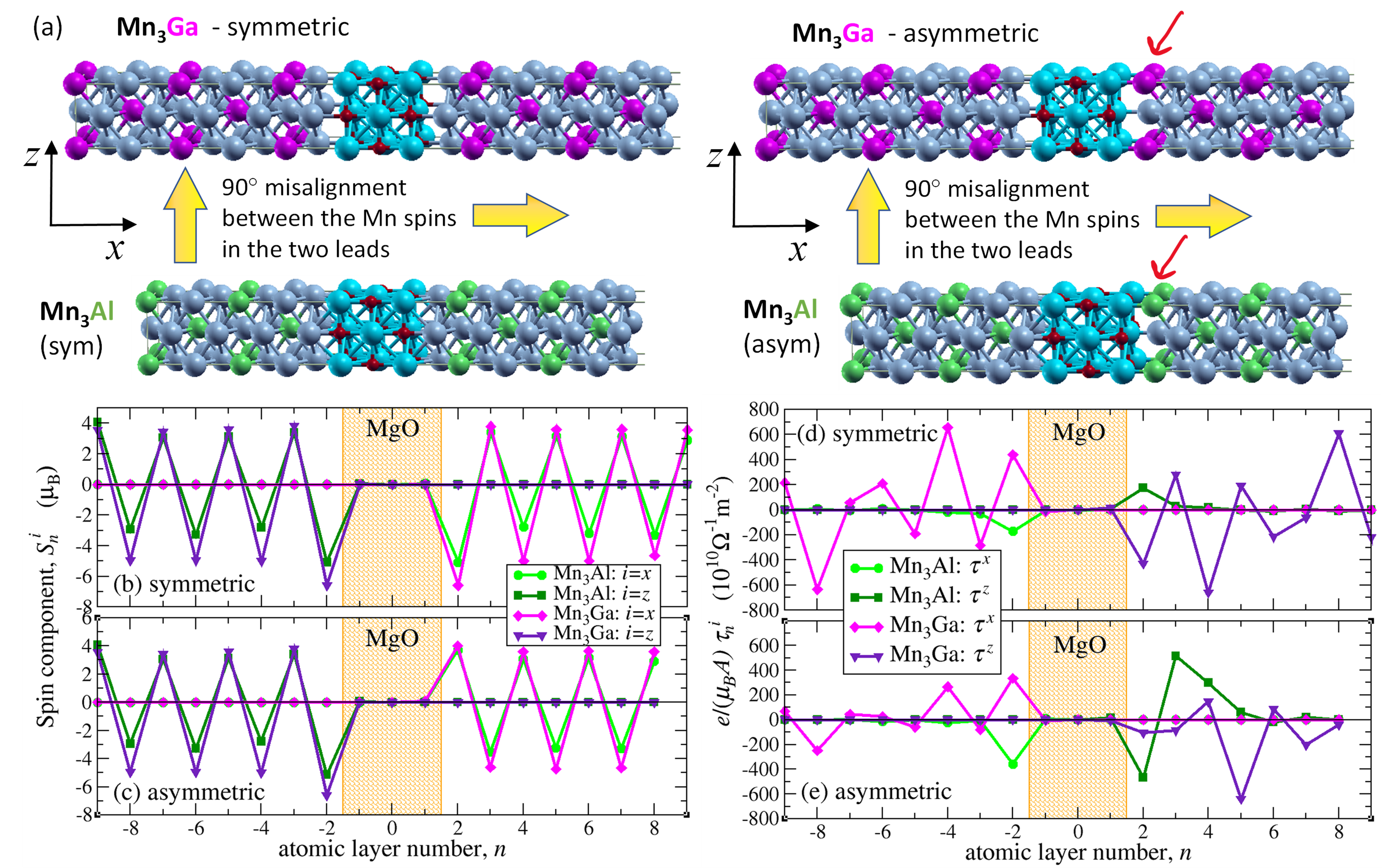}{(a) Schematic of the four MTJs considered, including the directions of the spin quantisation axes in the two leads at the 90\textdegree~alignment for the STT calculations. Note, that the corresponding 'asymmetric' MTJs have one layer of Mn removed from the right interface. (b,c) Self-consistently calculated layer-resolved spin-components ($x$ and $z$) for the four MTJs at equilibrium (see legend for the color code). (d,e) Corresponding layer-resolved in-plane STT components calculated at the Fermi level in linear response regime (so-called, torkance $\boldsymbol{\tau}=d\bm{T}(V)/dV$ at the limit $V=0$, where $\bm{T}(V)$ is the STT, see Ref. \onlinecite{Rungger2018} for precise definition) in the four different MTJs (same colour code). $A=a^2=16.81$~\AA$^2$ is the cross-sectional area of the junction (it is the same for all; note the periodic boundary conditions in the $x-y$ plane).}{fig02}

For Mn$_3$Ga we are using a longitudinal lattice constant $c=6.6$~\AA, which has been found to reproduce more reasonably the experimental values of the Mn spins (as extracted from neutron-diffraction results \cite{Rode2013}), within the LDA [see Fig. \ref{fig02}(b,c) for the computed layer-resolved magnetic moments], compared to the GGA-relaxed or the experimental lattice parameter values $c=7.1$~\AA~(larger by about 7~\%), for which the LDA-computed magnetic moments are larger by over 15~\% with respect to the $c=6.6$~\AA~case and outside the experimental range. We assume that, with these structural amendments (consistent also with Ref. \onlinecite{Stamenova2021} for Mn$_3$Ga-based MTJs), the LDA, which is not currently replaceable by GGA in our non-collinear-spin method for spin-transfer torque (STT)\cite{Stamenova2021}, captures the essential Fermi-surface properties of the two materials relevant for the linear-response regime investigated here and we continue exclusively with the LDA and the described above lattice parameters of both materials in Section \ref{Sect2}. 

Consequently, for the just described geometries, both compounds are ferrimagnetic with Mn$_3$Al being fully compensated (total magnetic moment $\simeq0$ ), while Mn$_3$Ga having a 2.6 $\mathrm{\mu_B}$ per cell (1.3~$\mathrm{\mu_B}$ per formula unit). The advantage of having a very low-spin-moment lead is to reduce the demagnetizing field, but more importantly, to reduce the critical current for a spin-transfer torque switching, which we will discuss it in the next Section \ref{Sect2}. We should mention here that the magneto-crystalline anisotropy calculated for Mn$_3$Ga is much bigger than that of Mn$_3$Al and this is because of the much stronger spin-orbit interaction at the Ga site compared to Al. The local spins (extracted by Mulliken population analysis) on the two Mn sublattices in Mn$_3$Al are nearly fully compensated with 3.31 $\mathrm{\mu_B}$ on the Mn$_\mathrm{I}$ site and -1.58 $\mathrm{\mu_B}$ on the Mn$_\mathrm{II}$ site, respectively. In Mn$_3$Ga the corresponding values are 3.53 and -2.46 $\mathrm{\mu_B}$ for Mn$_\mathrm{I}$ and Mn$_\mathrm{II}$, respectively, which are consistent with the measured moments\cite{Rode2013}.

\section{Spin-transport in \texorpdfstring{M\MakeLowercase{n}}{Mn}$_3$\texorpdfstring{A\MakeLowercase{l}}{Al} and \texorpdfstring{M\MakeLowercase{n}}{Mn}$_3$\texorpdfstring{G\MakeLowercase{a}}{Ga} junctions with \texorpdfstring{M\MakeLowercase{g}}{Mg}O barriers}
\label{Sect2}

Four different junctions based on Mn$_3$Al and Mn$_3$Ga all sandwiching 3 MLs of MgO have been investigated [Fig. \ref{fig02}a]. In the junctions, which we refer to as 'symmetric', the two interfaces are the same, that is, in both cases the interface is between the MgO and the Mn$_\mathrm{II}$-plane of the DO$_{22}$ lattice and overall the junction is mirror-symmetric with respect to the central plane in the MgO. In the 'asymmetric' junctions we have removed one monolayer of Mn from the right interface, but all distances, including the interface spacing, are preserved. Note that these geometries have not been relaxed -- the interface distance we have chosen in the Mn$_3$Al case is 2~\AA, while in the Mn$_3$Ga junctions we have chosen 2.2~\AA, as motivated in Ref. \onlinecite{Stamenova2021}. 

In Fig. \ref{fig02} we compare the linear-response STT \cite{Rungger2018, Stamenova2021} for a 90\textdegree~ misalignment of the spin polarisations in the two leads, computed within the non-equilibrium Green's function (NEGF) open-boundaries method implemented in the {\sc Smeagol} code~\cite{Alex04}. Panels (b) and (c) show the layer-resolved local moments in the scattering region -- in (b) the mirror symmetry is readily observed between the $x$ and $z$ components of the spins on both sides of the junction. In comparison, in (c) the asymmetry after the removal of one Mn plane at the right interface is evident. It is worth noting that the spins in the left lead across the MgO appear practically unaffected by this local structural disturbance on the right interface, in both junctions. 

The calculated layer-resolved in-plane STT for the symmetric junctions [Fig.\ref{fig02}(d)] displays a perfect left-to-right symmetry as well -- this time an inversion symmetry with respect to the central layer of MgO. For both ferrimagnets the STT at the opposite interfaces has an opposite sign. There is, however, a qualitative difference between the two ferrimagnets -- if the STT in the Mn$_3$Al case is localised at the interface, in Mn$_3$Ga it shows the familiar long-range oscillatory decay with periodicity determined by the difference of majority and minority spin wave-vectors of the $\Delta_1$-symmetry band in Mn$_3$Ga, as described in Ref. \onlinecite{Stamenova2021}. This ideal inversion symmetry of the STT across the barrier, however, will hinder the switching of the such ideally symmetric junctions. For instance, a positive $x$-component of STT will rotate anti-clockwise the Mn$_\mathrm{II}$ spins at the left interface (aligned along $-z$), and also anti-clockwise the Mn$_\mathrm{II}$ spins aligned along $-x$ at the right-hand-side interface. Hence such an alignment of the STT cannot drive switching of one lead with respect to the other. It is worth noting, however, that the STT changes sign in the next bi-layer -- this will act against the sub-lattice exchange coupling. This is would enable the excitation of high frequency anti-phase modes and potentially lead to fasted switching dynamics. It is likely that the ideal symmetry in our calculations will be broken in real structures and there will be an STT imbalance leading to the switching of one of the layers. In the Mn$_3$Al junction the STT has analogous symmetry, which does not lead to switching. It is, however, much more localised at the interfacial layer and significantly lower in amplitude. 

\myFig{1.5}{1.2}{true}{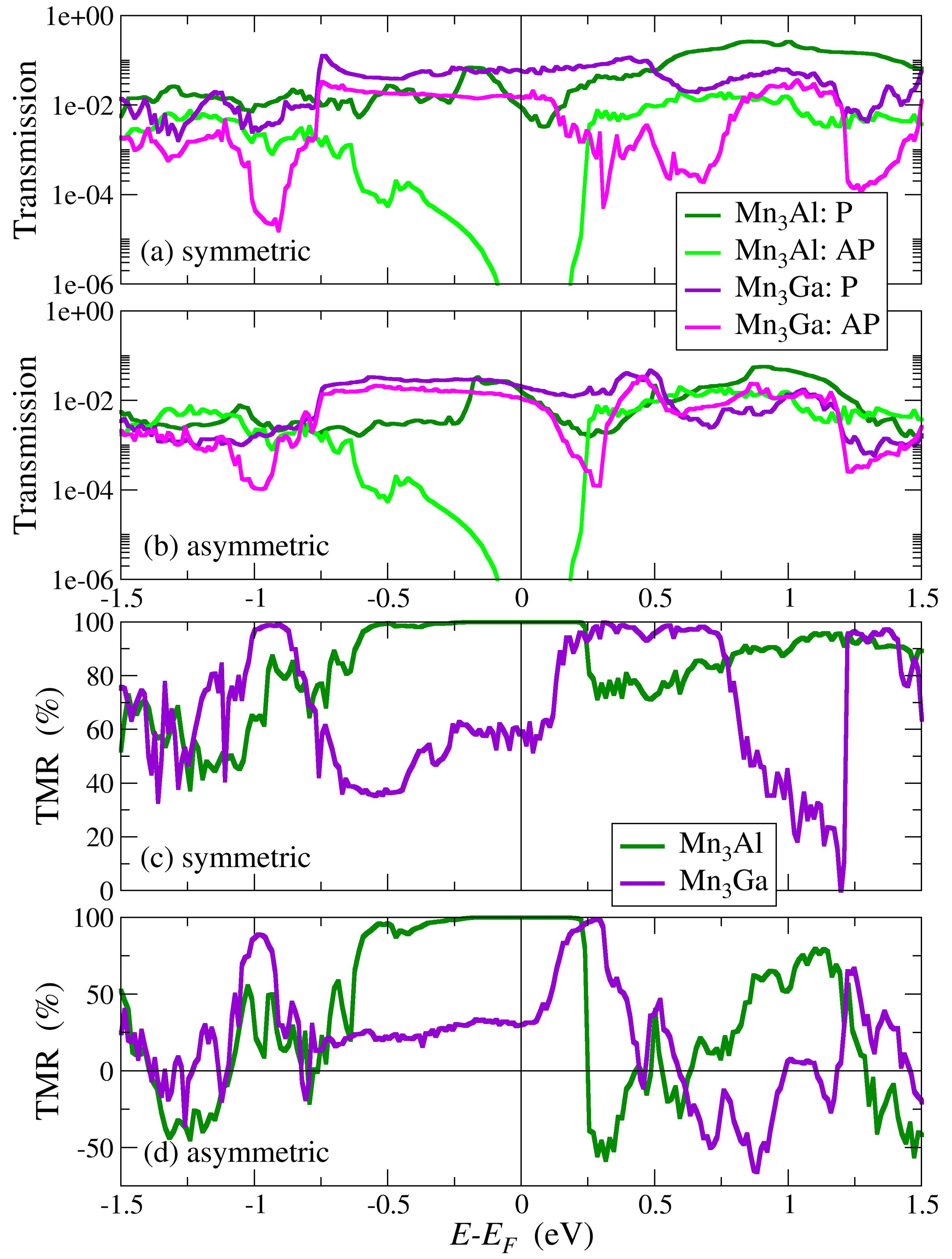}{Energy-resolved properties at 0~V equilibrium for both the P and AP states of the junctions (see text). (a,b) transmission coefficients and (c,d) TMR ratio, defined as $\mathrm{TMR}=(T_\mathrm{P}-T_\mathrm{AP})/(T_\mathrm{P}+T_\mathrm{AP})$ for the two symmetric and the two asymmetric MTJs, respectively.}{fig03}

To illustrate the effect of possible structural imperfection we consider the 'asymmetric' versions of both junctions -- with removed Mn-Mn monolayer from the right interface, while keeping the interface spacing unchanged [Fig. \ref{fig02}(a)]. In this case the STT profiles change substantially [Fig.\ref{fig02}(d,e)]. Now we see both in the Mn$_3$Al and Mn$_3$Ga junctions the STT becomes asymmetric on both sides of the barrier. This time it shows a tendency to rotate spins in opposite directions, especially in the Mn$_3$Al case, thus driving a switching of one of the layers with respect to the other. In the asymmetric Mn$_3$Al junction the STT is significantly increased in magnitude from the symmetric case, especially on the right-hand side, where we see a large STT also beyond the interfacial layer. 

We then examine the energy dependence of the total transmissions in the four junctions in their collinear spin states -- parallel (P) state in which the Mn spin in each sublattice (Mn$_\mathrm{I}$ or Mn$_\mathrm{II}$ type) on the two sides of the barrier are parallel, and AP state, when they are antiparallel. We clearly see the half-metallicity of Mn$_3$Al manifesting itself in the vanishing transmission in the AP state around the Fermi level. This in turn drives a large TMR (practically 100\% from the conservative definition with the sum of the transmission coefficients in the denominator) for a wide range of energies around $E_\mathrm{F}$ in both the symmetric and the asymmetric junctions [green curve in Fig. \ref{fig03} (c,d)]. In the case of Mn$_3$Ga we find a TMR of about 60~\% for the symmetric junction, which drops to 30~\% for the asymmetric case. We find a significant enhancement of the TMR for energies higher than 0.2~eV above $E_\mathrm{F}$, which is due to the $\Delta_1$ symmetry band edge and the band gap opening above that energy for majority spin in the $\Gamma-Z$ (transport) direction \cite{Stamenova2021} (note that the position of this band edge above $E_\mathrm{F}$ is consistent between the LDA and GGA calculations in Fig.\ref{fig01}). The role of the interfacial asymmetry appears to be in reducing the transmission in the P state, where in the asymmetric junctions the interfacial spins are pointing in opposite directions, giving rise to additional scattering at the interfaces.  

\myFigWide{1.7}{1}{true}{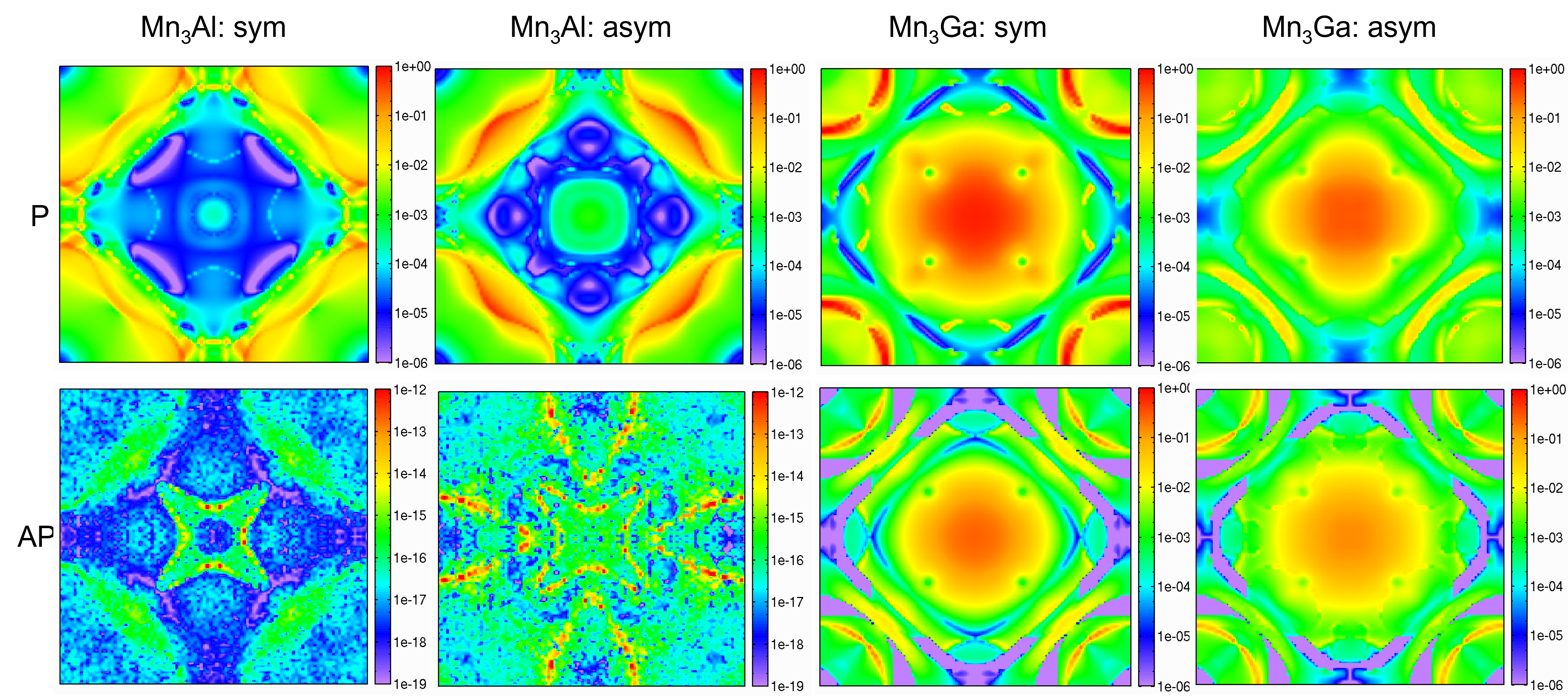}{Contour portraits of the total transmission coefficient (both spin species) at the Fermi level decomposed over the transverse 2D-BZ [as a function of $(k_x,k_y)$ on each panel, where $-\pi/a\leq k_{x,y}\leq\pi/a$]. From left to right, the four different MTJs are depicted (as indicated above), while the two rows correspond to the parallel (P) and anti-parallel (AP) spin states of the junctions, respectively (see text). Color-code bars for each panel are for the $T(k_x,k_y,E=E_\mathrm{F})$, i.e. dimensionless transmission probabilities (see, e.g. Ref.\onlinecite{Rungger2018}).}{fig04}

This can also be seen in the 2D portraits of the Fermi-level transmission coefficients in the transverse 2D Brillouin zone (2D-BZ) (Fig. \ref{fig04}). In the Mn$_3$Ga case the transmission in all spin-states and junction geometries is predominantly around the $\Gamma$ point. The P states show stronger transmission in both junctions, while in the asymmetric case both spin-polarised transmissions are somewhat suppressed with respect to the corresponding ones in the symmetric MTJ. In contrast to Mn$_3$Ga, the transmission around the $\Gamma$ point is suppressed in the two Mn$_3$Al junctions in the P state, as in this material there is a band gap along $\Gamma-Z$ direction for majority (up) spins [see Fig. \ref{fig01}(a)]. Interestingly, despite the small quantitative changes, which the asymmetric interface introduces in some regions of the transverse 2D-BZ, for the small range of wavevector angles relevant to realistic specular transport around the $\Gamma$ point, the transmission appears relatively unaffected by the interfacial detail. This effect is expected to be enhanced further for thicker MgO barriers, where the transmission is more strongly filtered in the vicinity of $\Gamma$ point and we expect the overall transmission in Mn$_3$Al-based MTJs to be even less sensitive to features of the interfaces. We also note that in the AP state of the Mn$_3$Al junctions the transmission is extremely small and at the verge of computational accuracy, as expected from its halfmetallicity (see Fig. \ref{fig01}).       

\section{Conclusions}

We have compared linear-response spin-polarised transport properties in ferrimagnetic tunnel junctions made from two similar Mn-based binary Heuslers in their tetragonal phase sandwiching a MgO barrier. Junctions featuring Mn$_3$Al offer a different compromise between the magnitude of STT and TMR, when compared to otherwise equivalent Mn$_3$Ga-based ones. The overall transmission, and therefore the expected differential conductivity, is higher in the ferrimagnetic Mn$_3$Ga case, together with the low-bias STT. The absence of one type of spin states for Mn$_3$Al-based junctions, leads to smaller overall absorbed momentum (with STT acting only within a couple of lattice spacings away from the interfaces) and a significantly suppressed low-bias differential conductance. It can be argued, therefore that at least at low temperature and applied bias, when compared on the basis of STT per unit dissipated power, the Mn$_3$Ga-based junctions would offer an edge over the otherwise better in their TMR performance Mn$_3$Al counterparts. The relatively low sensitivity of the calculated parallel-state transmission (normal to the interface) and also of the STT, although to a lesser extent, on the exact surface reconstruction in both Mn$_3$Al and Mn$_3$Ga-based junctions, may further motivate the experimental verification of spectroscopic TMR and STT features (at low temperatures), hence, finite-bias calculations of the same would be well-warrantied. In view of the above, aluminium-rich solid solutions of Mn$_3$Al-Mn$_3$Ga, with or without Sn doping, could offer enhanced low-bias TMR performance, while preserving tetragonality, at manageable growth-induced strain, in practical device stacks.

\begin{acknowledgments}
We gratefully acknowledge funding from the Science Foundation Ireland (SFI Grant No. 18/SIRG/5515 and 18/NSFC/MANIAC). We thank the Irish Centre for High-End Computing (ICHEC) and the Trinity Research IT Centre (TCHPC) for the provision of computational facilities and support.

\end{acknowledgments}

Copyright (2022) Authors. This article is distributed under a Creative Commons Attribution (CC BY) License.

\nocite{*}
\bibliography{main}% Produces the bibliography via BibTeX.

\end{document}